\documentclass[12pt]{article}

\usepackage{amsmath,amssymb,amsbsy,amsfonts,latexsym,graphicx}
\usepackage{color,array}

\numberwithin{equation}{section}

\allowdisplaybreaks

\evensidemargin \oddsidemargin

\begin{document}


\begin{titlepage}

\renewcommand{\thefootnote}{\fnsymbol{footnote}}


\begin{flushright}
{\tt KIAS-P12043}
\end{flushright}

\vspace{15mm}
\baselineskip 9mm
\begin{center}
{\Large \bf Consistent bilinear Wess-Zumino term \\
  for open AdS superstring}
\end{center}

\baselineskip 6mm
\vspace{10mm}
\begin{center}
Ee Chang-Young$^{a,b}$\footnote{\tt cylee@sejong.ac.kr},
Hiroaki Nakajima$^{b,c}$\footnote{{\tt nakajima@kias.re.kr}, 
moving to National Taiwan University from August.}, and
Hyeonjoon Shin$^d$\footnote{\tt hyeonjoon@postech.ac.kr}
\\[10mm]

{\sl $^a$Department of Physics, Sejong University, Seoul 143-747, Korea}
\\[3mm]
{\sl $^b$School of Physics, Korea Institute for Advanced Study,\\
207-43 Cheongnyangni-dong, Dongdaemun-gu,
Seoul 130-722, Korea}
\\[3mm]
{\sl $^c$Department of Physics, Kyungpook National University,\\
Taegu 702-701, Korea}
\\[3mm]
{\sl $^d$Department of Physics, Pohang University of Science
  and Technology,\\
  and Asia Pacific Center for Theoretical Physics, \\
  Pohang, Gyeongbuk 790-784, Korea }
\end{center}

\thispagestyle{empty}

\vfill
\begin{center}
{\bf Abstract}
\end{center}
\noindent
We consider the open superstring action in AdS$_5 {}\times{}$S$^5$
background with the bilinear Wess-Zumino term, which has been
modified in such a way that a certain total derivative term is absent,
and give the covariant description of supersymmetric D-branes.  We show
that the modification of bilinear Wess-Zumino term is necessary to
describe correctly the 1/2-BPS D(-1)-brane in AdS$_5 {}\times{}$S$^5$
background, while the classification of
other supersymmetric D-branes does not depend on such modification.
\\ [5mm] Keywords : Wess-Zumino term, D-brane, $\kappa$-symmetry

\vspace{5mm}
\end{titlepage}

\baselineskip 6.6mm
\renewcommand{\thefootnote}{\arabic{footnote}}
\setcounter{footnote}{0}

\section{Introduction}

The type IIB supertring in the AdS$_5 {}\times{}$S$^5$ background is
an important ingredient in the study of AdS/CFT correspondence
\cite{AdS/CFT,GKPW}.  As is well known, it has been
described in \cite{Metsaev:1998it,Kallosh:1998zx} as a supersymmetric
Green-Schwarz type sigma model based on the fact that the
AdS$_5 {}\times{}$S$^5$ background has the coset superspace structure.
After the construction of its action, it has been observed
in a similar construction in other AdS type background
\cite{Berkovits:1999zq} that
the symmetry superalgebra corresponding to the coset superspace
has the $\mathbf{Z}_4$-automorphism.  This has motivated an
alternative description of type IIB superstring in the
AdS$_5 {}\times{}$S$^5$ background \cite{Roiban:2000yy}, which has provided
a basis for the study of integrability in the superstring
theory \cite{Bena:2003wd}.\footnote{For a comprehensive review on the
development based on the alternative formulation of type IIB superstring
in AdS$_5 {}\times{}$S$^5$, see \cite{Arutyunov:2009ga} for example.}

The superficial difference between two descriptions is in the
Wess-Zumino (WZ) term.  In the conventional formulation by Metsaev and
 Tseytlin \cite{Metsaev:1998it}, the WZ term is given by
\begin{align}
-2i \int^1_0 dt  \int_{\Sigma} \widehat{L}^A \wedge
 \bar{\theta}^I \Gamma_A \tau_3^{IJ} \widehat{L}^J \,,
\label{mtwz}
\end{align}
where $\Sigma$ denotes the string worldsheet, and $\widehat{L}^A$
($\widehat{L}^I$) is the vetor (spinor) superfield or the Maurer-Cartan
one-form superfield $L^A$ ($L^I$) with the rescaling
$\theta \rightarrow t \theta$, that is,
$\widehat{L}^A(X, \theta) = L^A(X, t\theta)$
($\widehat{L}^I(X, \theta) = L^I(X, t\theta)$).  (All the detailed
expressions for the superfields including our notation and convention
is given in the appendix.)  From (\ref{mtwz}), we see that there is an
integration in terms of an auxiliary parameter $t$.
Contrary to this, the alternative formulation of \cite{Roiban:2000yy}
states that the WZ term does not involve such auxiliary integration and
furthermore 
is manifestly bilinear 
with respect to the superfield, which is written as
\begin{align}
\int_\Sigma \bar{L}^I \wedge \Gamma_* \tau_1^{IJ} L^J
\label{rswz}
\end{align}
in the 32 component notation.\footnote{By using
$\Gamma_* \equiv i\Gamma_{01234}$ of (\ref{gdef}), and the `5+5'
way \cite{Metsaev:1998it} of splitting the Dirac gamma
matrices, $\Gamma^a = \gamma^a \times 1 \times \sigma_1$,
$\Gamma^{a'} = 1 \times \gamma^{a'} \times \sigma_2$,
($a=0,\dots,4$, $a'=5,\dots,9$) where $\sigma_k$
are Pauli matrices, the integrand reduces to
$\bar{L}^I \wedge \tau_1^{IJ} L^J$, which is the usual form considered
in the literature.}
It is obvious that the WZ term  of (\ref{rswz}) is simpler than that of
(\ref{mtwz}) and hence seems to be advantageous in the study of
superstring theory.  Actually, it has been demonstrated that the bilinear
WZ term is more practical to explore the algebraic or the dynamical aspect
of type IIB superstring in the AdS$_5 {}\times{}$S$^5$ background
\cite{Hatsuda:2000mn,Hatsuda:2001pp,Hatsuda:2001xf}.
We would like to note that the same kind of bilinear WZ term arises also
from the study of AdS/CFT correspondence for the non-critical
strings\cite{Polyakov:2004br}.

One peculiar point in the structure of (\ref{rswz}) is that there
is a total derivative term, $d\bar{\theta}^I \wedge \Gamma_*
\tau_1^{IJ} d\theta^J$, which is given as the leading order term
when we expand the integrand in terms of the fermionic coordinate
$\theta$ by using the expression of $L^I$ given in
(\ref{superfld}).  It has been pointed out in
\cite{Hatsuda:2002hz} that such a term should be subtracted from
the WZ term to have the correct charge for the massive string
excitations\footnote{The charge corresponds to the string winding
number in the flat space limit \cite{Hatsuda:2002hz}.  So, unless
any direction is somehow compactified, it seems that the physics
does not depend on the total derivative term.} and thus the
bilinear WZ term of (\ref{rswz}) should be modified as
\begin{align}
S_{WZ} = \int_\Sigma  (\bar{L}^I \wedge \Gamma_* \tau_1^{IJ} L^J
             - d\bar{\theta}^I \wedge \Gamma_* \tau_1^{IJ} d\theta^J ) \,.
\label{wz}
\end{align}

The modification (\ref{wz}) has been proposed to deal with the
problem in the closed string case.  Now, one may be interested in
the open string case and ask what the effect of the subtraction of
the total derivative term is. In this paper, we address this question
by considering the open string description of D-branes in the
covariant setting. Firstly, following the prescription of
\cite{Bain:2002tq}, we consider the $\kappa$-symmetry variation of
the superstring action.  In order to make the action to be
$\kappa$-symmetric, we impose suitable open string boundary
conditions on the worldsheet boundary.  In this way, we give the
covariant description of D-branes and classify the supersymmetric
1/2-BPS D-branes in the AdS$_5 {}\times{}$S$^5$ background. We then
compare our result with that obtained in different contexts
\cite{Skenderis:2002vf,Sakaguchi:2003py}.  As we will see, the
comparison shows that it is necessary for the bilinear WZ term
(\ref{rswz}) to be modified as (\ref{wz}) for the full correct
classification of 1/2-BPS D-branes. More precisely, the modified
bilinear WZ term (\ref{wz}) should be used for the description of
1/2-BPS D(-1)-brane, that is, D-instanton. On the other hand, for
other D$p$-branes with $p \ge 1$, we do not need to care about the
presence of the total derivative term.

This paper is organized as follows.  The covariant open string
description of 1/2-BPS D-branes is given in the next section. In
Sec.~\ref{allorder}, it is shown that the result of
Sec.~\ref{covd} is valid at full orders in the fermionic
coordinate $\theta$. The final section is devoted to our
conclusion.  In Appendix, we give the expressions for the
superfields together with the notations and conventions.

\section{Covariant description of D-branes}
\label{covd}

In the original proposal for the covariant description of D-branes by
using the Green-Schwarz open superstring action \cite{Lambert:1999id},
an arbitrary variation of the action is considered and
suitable open string boundary conditions for making the action
invariant under the variation are investigated. However,
as noted in \cite{Bain:2002tq}, the very $\kappa$-symmetry variation,
not arbitrary one, is enough at least for the description of
supersymmetric D-branes, because the $\kappa$-symmetry leads to the
matching of dynamical degrees of freedom for bosons and fermions on
the worldsheet and hence ensures the object described by the open string
supersymmetric.\footnote{The suggestion using the $\kappa$-symmetry
\cite{Bain:2002tq} has been demonstrated in the type IIB pp-wave
background.  In subsequent works, it has been successfully applied to
other string theory backgrounds \cite{Sakaguchi:2003py,Hyun:2002xe}.}
In this section, we investigate the open string
boundary conditions under which the superstring action with the
modified bilinear WZ term of (\ref{wz}) is $\kappa$-symmetric.

The $\kappa$-symmetry transformation rules in superspace are given by
\begin{align}
\delta_\kappa Z^M L_M^A = 0 \,, \quad
\delta_\kappa Z^M L_M^I = \eta^I \,, \quad
\eta^I \equiv (\delta^{IJ}+\tau_3^{IJ} \Gamma) \kappa^J \,,
\label{k-rule}
\end{align}
where $\kappa^I$ is the $\kappa$-symmetry transformation parameter and
$\Gamma$ is basically the pullback of $\Gamma_{AB}$ onto the string
worldsheet with the properties, $\Gamma^2 =1$ and $\text{Tr}\Gamma=0$,
whose detailed expression is not needed here.  Since the bulk part
of the superstring action is $\kappa$-symmetric by construction, what
we have under the $\kappa$-symmetry variation are the boundary
contributions.  It should be noted here that, as shown in
\cite{Bain:2002tq}, the kinetic part of the superstring action does not
give any boundary contribution due to $\delta_\kappa Z^M L_M^A = 0$
in (\ref{k-rule}).  Thus we can focus only on the WZ term rather
than the full superstring action.

The WZ term has an expansion in terms of the
fermionic coordinate $\theta$ up to the order of $\theta^{32}$.  Although
it is so, we will consider the expansion only up to quartic order in
$\theta$ in this section.  As we will see in the next section, all the
nontrivial information for the description of D-branes is obtained already
from the terms in such restricted expansion.
Then the WZ action (\ref{wz}) expanded up to the desired order is
written as
\begin{align}
S_{WZ} = S^0 + S^{\text{spin}} + S^{\mathcal{M}^2} + \dots \,,
\end{align}
where the dots represent the higher order terms and we have divided
the terms of our interest into three parts, that is,
$\mathcal{M}^2$ dependent part $S^{\mathcal{M}^2}$ 
(see Appendix for the definition of $\mathcal{M}^{2}$. ),
the spin connection dependent part $S^{\text{spin}}$, and
the part $S^0$ containing the remaining terms.
These three parts have the following expressions.
\begin{align}
S^0 &= \int_\Sigma
   \left(
       i e^A \wedge \bar{\theta}^I \Gamma_A \tau_3^{IJ} d\theta^J
     + \frac{1}{4} e^A \wedge e^B \bar{\theta}^I \Gamma_A
             \Gamma_* \Gamma_B \tau_1^{IJ} \theta^J
   \right) \,,
\notag \\
S^{\text{spin}} &= - \int_\Sigma \omega^{AB} \wedge
   \left(
     \frac{1}{2}  \bar{\theta}^I \Gamma_{AB}
          \Gamma_* \tau_1^{IJ} d\theta^J
    + \frac{1}{4^2} \omega^{CD}
           \bar{\theta}^I \Gamma_{AB} \Gamma_* \Gamma_{CD}
           \tau_1^{IJ} \theta^J
    + \frac{i}{4}  e^C
           \bar{\theta}^I \Gamma_{ABC} \tau_3^{IJ} \theta^J
    \right) \,,
\notag \\
 S^{\mathcal{M}^2} &=  \frac{1}{3} \int_\Sigma
    \overline{D\theta^I} \wedge \Gamma_* \tau_1^{IJ}
    (\mathcal{M}^2)^{IJ} D \theta^J  \,.
\label{3parts}
\end{align}

For the variation of these
parts, it is now convenient to express the variation $\delta_\kappa X^\mu$
in terms of $\delta_\kappa \theta^I$ by using the transformation
rule $\delta_\kappa Z^M L_M^A = 0$ of (\ref{k-rule}) as follows:
\begin{align}
\delta_\kappa X^\mu = - i \bar{\theta}^I \Gamma^\mu
      \delta_\kappa \theta^I + \mathcal{O} (\theta^4) \,,
\end{align}
where $\mathcal{O} (\theta^4)$ leads to the terms of higher order
than quartic one in the resulting variation of the WZ term and thus
is not of our concern here.  By utilizing this, we first consider the
boundary contributions from the $\kappa$-symmetry variation of $S^0$
found as
\begin{align}
\delta_\kappa S^0
= \int_{\partial \Sigma} &
  \bigg[ -i dX^\mu e_\mu^A (\bar{\theta}^I \Gamma_A \tau_3^{IJ}
                       \delta_\kappa \theta^J)
    + (\bar{\theta}^I \Gamma_A \tau_3^{IJ} d\theta^J)
      (\bar{\theta}^K \Gamma^A \delta_\kappa \theta^K)
\notag \\
& -\frac{i}{2} dX^\mu e_\mu^B
   (\bar{\theta}^I \Gamma_A \Gamma_* \Gamma_B \tau_1^{IJ} \theta^J)
   (\bar{\theta}^K \Gamma^A \delta_\kappa \theta^K) \bigg] \,,
\label{vars0}
\end{align}
where $\partial \Sigma$ represents the boundary of open string worldsheet.
We have three non-vanishing terms on the right hand side.  In order to
have the $\kappa$ invariance, they should vanish under a suitable set
of open string boundary conditions.  As for the first term, because
\begin{align}
d X^A \equiv dX^\mu e^A_\mu = 0 \quad (A \in D) \,,
\label{vc1}
\end{align}
where $A \in D~(N)$ means that $A$ is a direction of Dirichlet
(Neumann) boundary condition,
\begin{align}
\bar{\theta}^I \Gamma_A \tau_3^{IJ} \delta_\kappa \theta^J
\label{f1}
\end{align}
should vanish for $A \in N$.  For satisfying this, we impose the
following 1/2-BPS boundary condition
\begin{align}
\theta^I = P^{IJ} \theta^J
\label{bc}
\end{align}
with
\begin{align}
P^{IJ} =
  \left\{
     \begin{array}{ll}
         s \Gamma^{A_1\dots A_{p+1}} \tau_1^{IJ} &  (p=1~\text{mod}~4)\\
         s \Gamma^{A_1\dots A_{p+1}} \epsilon^{IJ} & (p=3~\text{mod}~4)
     \end{array}
  \right. \,,
\label{pmat}
\end{align}
where all the indices $A_1,\dots, A_{p+1}$ are those for Neumann
directions and
\begin{align}
s = \left\{
      \begin{array}{ll}
         1 & \text{for}~  X^0 \in N \\
         i & \text{for}~  X^0 \in D
      \end{array}
    \right. \,.
\end{align}
We note that $p$ should be odd because $\theta^1$ and $\theta^2$ have the
same chirality and, for any odd $p$,
\begin{align}
P^{IJ}P^{JK}=\delta^{IK} \,, \quad
\bar{\theta}^I = - \bar{\theta}^J P^{JI} \,.
\label{bcprop}
\end{align}
Without much difficulty, we can now check that the boundary condition
(\ref{bc}) makes the term of (\ref{f1}) vanish explicitly, that is,
$\bar{\theta}^I \Gamma_A \tau_3^{IJ} \delta_\kappa \theta^J = 0$ for
$A \in N$.  In turn, this result for the first term of (\ref{vars0})
immediately leads us to have the vanishing condition for the second
term as
\begin{align}
\bar{\theta}^I \Gamma^A \delta_\kappa \theta^I = 0 \quad (A \in D) \,.
\label{vc2}
\end{align}
The boundary condition (\ref{bc}) can be imposed again to show that this is
indeed the case.  So, up to this point, all odd $p$, that is, D$p$-branes
with $p= -1,1,3,5,7,9$ are possible.

The situation changes at the third term of (\ref{vars0}).  From Eqs.
(\ref{vc1}) and (\ref{vc2}), we see that the vanishing condition of
the term is
\begin{align}
\bar{\theta}^I \Gamma_A \Gamma_* \Gamma_B \tau_1^{IJ} \theta^J = 0 \quad
(A, B \in N) \,.
\label{vc3}
\end{align}
Due to the presence of $\Gamma_*$, there are restrictions in the
number of Neumann directions in AdS$_5$ or $S^5$ for satisfying this
condition.  Let us denote $n$ ($n'$) as the number of Neumann directions
among $0,\dots,4$ ($5,\dots,9$).  Then we have the relation,
\begin{align}
n+n' = p+ 1 \,,
\end{align}
which means that both of $n$ and $n'$ are even or odd because $p+1$ is
even.  Simple calculation shows that the condition (\ref{vc3}) is
satisfied for the following cases:
\begin{align}
\begin{array}{ll}
n,~n' : \mbox{even} & \quad (p=1~\mbox{mod}~4) \\
n,~n' : \mbox{odd} & \quad (p=3~\mbox{mod}~4) \,.
\end{array}
\label{susybc}
\end{align}
This gives us the information about the directions to which a
1/2-BPS D-brane can extend and shows that D9-brane is not 1/2-BPS.
We would like to note here that the
D(-1)-brane, that is, D-instanton is excluded in this restriction,
since $n=n'=0$ is not allowed for $p=3$ mod $4$.  However, the
third term vanishes even for the D-instanton case, because $dX^A = 0$
for all $A$ for the D-instanton boundary condition.  Thus, the
D-instanton is also one of the 1/2-BPS D-branes.

We now turn to the boundary contributions from the $\kappa$-symmetry
variation of $S^{\text{spin}}$, which are found as
\begin{align}
\delta_\kappa S^{\text{spin}}
= \frac{1}{2} \int_{\partial \Sigma} &
  \bigg[
     (\bar{\theta}^I \Gamma_{AB} \Gamma_* \tau_1^{IJ}
          \delta_\kappa \theta^J) dX^\mu
   + \frac{1}{2} (\bar{\theta}^I \Gamma_C \Gamma_{AB} \tau_3^{IJ} \theta^J)
                 (\bar{\theta}^K \Gamma^C \delta_\kappa \theta^K) dX^\mu
\notag \\
& + i (\bar{\theta}^I \Gamma_{AB} \Gamma_* \tau_1^{IJ} D \theta^J)
      (\bar{\theta}^K \Gamma^\mu \delta_\kappa \theta^K)
  \bigg] \omega_\mu^{AB} \,.
\label{varsspin}
\end{align}
As for the first two terms on the right hand side, 
by repeating the same procedure 
applied to $\delta_\kappa S^0$ of (\ref{vars0}), the boundary
condition (\ref{bc}) with the restriction
(\ref{susybc}) leads us to have some spinor bilinears which vanish
at the boundary,
\begin{gather}
\bar{\theta}^I \Gamma_{AB} \Gamma_* \tau_1^{IJ}
\delta_\kappa \theta^J = 0 \quad (A,B \in N(D)) \,,
\notag \\
\bar{\theta}^I \Gamma_C \Gamma_{AB}  \tau_3^{IJ} \theta^J = 0 \quad
(A,B \in N (D), C \in N) \,.
\end{gather}
Although these eliminate the boundary contributions with the corresponding
index structure, other contributions do not vanish.  By the way,
one common property of those surviving contributions is that they
are proportional to the spin connection $\omega^{AB}$ with
$A \in N$ and $B \in D$ (or $A \in D$ and $B \in N$).  From the
expression of the spin connection (\ref{zehn}), we see that those
contributions vanish if the Dirichlet directions are set to zero.
This means that a given D-brane is 1/2-BPS if it is placed at the
coordinate origin in its transverse directions.

For the consideration of the third therm on the right hand side of
(\ref{varsspin}), the following relation at the boundary is useful.
\begin{align}
(D P \theta)^I = (P D \theta)^I
             - \omega^{AB} \Gamma_{AB} P^{IJ} \theta^J \,,
\label{diffcov}
\end{align}
where $A \in N$ and $B \in D$.  By utilizing this, we can show that
all the non-vanishing boundary contributions are proportional to
$\omega^{AB}$ with  $A \in N$ and $B \in D$ (or $A \in D$ and $B \in N$)
like the case of first two terms of (\ref{varsspin}), and hence  vanish
at the origin of the Dirichlet directions and we have
the $\kappa$-invariance.
As a remark, we would
like to note that the D-instanton is actually exceptional because the
whole boundary contributions from $\delta_\kappa S^{\text{spin}}$ vanish
basically because of (\ref{vc1}) and (\ref{vc2}).  Thus, the D-instanton
is 1/2-BPS in every position.

Finally, there are boundary contributions from
$\delta_\kappa S^{{\mathcal M}^2}$ which are obtained as
\begin{align}
\delta_\kappa S^{{\mathcal M}^2}
= \frac{1}{6} \int_{\partial \Sigma} &
  \Big[
     2(\bar{\theta}^I \Gamma_A \delta_\kappa \theta^I)
     (\bar{\theta}^J \Gamma^A \tau_3^{JK} D\theta^K)
   - 2(\bar{\theta}^I \Gamma_A D\theta^I)
     (\bar{\theta}^J \Gamma^A \tau_3^{JK} \delta_\kappa \theta^K)
\notag \\
&  + (\bar{\theta}^I \Gamma_{ab} \Gamma_* \epsilon^{IJ}
            \delta_\kappa\theta^J)
     (\bar{\theta}^K \Gamma^{ab} \Gamma_* \tau_1^{KL} D\theta^L)
\notag \\
&  - (\bar{\theta}^I \Gamma_{ab} \Gamma_* \epsilon^{IJ} D\theta^J)
     (\bar{\theta}^K \Gamma^{ab} \Gamma_* \tau_1^{KL}
            \delta_\kappa\theta^L)
\notag \\
&  - (\bar{\theta}^I \Gamma_{a'b'} \Gamma_*' \epsilon^{IJ}
            \delta_\kappa\theta^J)
     (\bar{\theta}^K \Gamma^{a'b'} \Gamma_* \tau_1^{KL} D\theta^L)
\notag \\
&  + (\bar{\theta}^I \Gamma_{a'b'} \Gamma_*' \epsilon^{IJ} D\theta^J)
     (\bar{\theta}^K \Gamma^{a'b'} \Gamma_* \tau_1^{KL}
            \delta_\kappa\theta^L)
   \Big] \,.
\end{align}
We have checked that the boundary contributions vanish without any
additional condition.  However, we do not give any detailed
explanation, because the term containing ${\mathcal M}^2$ as well
as the terms of higher powers of ${\mathcal M}^2$ will be dealt
with all at once in the next section. So, we complete the
investigation of the open string boundary condition for the
$\kappa$-symmetry of the action expanded up to quartic order in
$\theta$, and hence the classification of 1/2-BPS D-branes, which
is summarized in the table \ref{btable}.  We note that our D-brane
classification is in complete agreement with that from the probe
analysis \cite{Skenderis:2002vf} and from the analysis using the
WZ term constructed by Metsaev and Tseytlin (\ref{mtwz})
\cite{Sakaguchi:2003py}.

\begin{table}
\begin{center}
\begin{tabular}{|c|c|}
\hline
D$p$  & $(n,n')$ \\
\hline \hline
D(-1) & $(0,0)$ \\
\hline
D1    & $(2,0)$ $(0,2)$ \\
\hline
D3    & $(1,3)$ $(3,1)$ \\
\hline
D5    & $(2,4)$ $(4,2)$ \\
\hline
D7    & $(3,5)$ $(5,3)$ \\
\hline
D9    & absent \\
\hline
\end{tabular}
\end{center}
\caption{1/2-BPS D-branes in the AdS$_5 {}\times{}$S$^5$ background.
$n$ ($n'$) is the number of Neumann directions in AdS$_5$ ($S^5$).}
\label{btable}
\end{table}

In our study, we have taken the modified bilinear WZ term (\ref{wz})
which does not include the total derivative term.
Before going to the next section, we consider the problem as to
whether the classification of 1/2-BPS D-branes is valid even if
such total derivative term is included in the WZ term.  Under the
$\kappa$-symmetry variation, we have
\begin{align}
\delta_\kappa \int_\Sigma d \bar{\theta}^I
        \wedge \Gamma_* \tau_1^{IJ} d\theta^J
=  2 \int_{\partial \Sigma} \delta_\kappa \bar{\theta}^I \Gamma_*
            \tau_1^{IJ}  d \theta^J \,,
\label{ktotal}
\end{align}
where there are only boundary contributions because the variation
of total derivative term does not give bulk contribution.  It is
not difficult to show that the boundary condition (\ref{bc}) with
the restriction (\ref{susybc}) makes the boundary contributions
vanish and ensures the $\kappa$-invariance.  On the other hand, we
find that the boundary condition for the D-instanton, which is not
in the criterion of (\ref{susybc}), cannot eliminate the right
hand side of (\ref{ktotal}).  Thus, the inclusion of the total
derivative term does not lead to the 1/2-BPS D-instanton while it
does not affect the classification of 1/2-BPS D$p$-branes with $p
\ge 1$.\footnote{Related to our result, it has been reported that
the D-instanton is distinguished from other D-branes also in the
construction of D-brane action in AdS space
\cite{Hatsuda:2004vi}.} This means that we should use the modified
bilinear WZ term (\ref{wz}) for the full correct classification
of 1/2-BPS D-branes.

\section{Validity at full orders in $\theta$}
\label{allorder}

We have shown that the boundary condition (\ref{bc}) with the
restriction (\ref{susybc}) makes the boundary contributions from
the $\kappa$-symmetry variation of the WZ term vanish up to the
quartic order in $\theta$.  In this section, we provide a proof that
such boundary condition is sufficient for showing the boundary
$\kappa$-symmetry of the WZ term even at higher orders in $\theta$
without any extra boundary condition, and thus the classification of
1/2-BPS D-branes in the AdS$_5 {}\times{}$S$^5$ background
summarized in the table \ref{btable} is valid at full orders in $\theta$.

For the investigation of $\kappa$-symmetry at higher orders in
$\theta$, it is not necessary to take the modified bilinear WZ term
(\ref{wz}) since the subtracted total derivative term is the leading order
one.  Thus it is enough to consider the unmodified bilinear WZ term
of (\ref{rswz}).
Then the boundary contribution from the $\kappa$-symmetry variation of
this term is obtained as
\begin{align}
2\int_{\partial \Sigma} \bar{\eta}^I \Gamma_* \tau_1^{IJ} L^J \,,
\label{kvar}
\end{align}
where we have used the $\kappa$-symmetry transformation rule
(\ref{k-rule}).  In order to proceed, we need to know the boundary
condition of $\eta^I$.  Since we can see that
$\delta_\kappa \theta^I = \eta^I + \mathcal{O}(\theta^2)$
from the transformation rule (\ref{k-rule}), it is natural to expect
that the boundary condition of $\eta^I$ is the same with that of
$\theta^I$, that is, $\eta^I = P^{IJ} \eta^J$
(or $\bar{\eta}^I = - \bar{\eta}^J P^{JI}$).  Although this seems a
naive expectation, it has been shown rigorously in \cite{Sakaguchi:2004}
that this is indeed the case.\footnote{For the rigorous proof, consult
the procedure of proving Eq.~(3.4) of \cite{Sakaguchi:2004}.}  If we
now impose this boundary condition in (\ref{kvar}) and carry out
a bit of manipulation with the condition (\ref{susybc}), then we have
\begin{align}
-2\int_{\partial \Sigma} \bar{\eta}^I \Gamma_* \tau_1^{IJ} P^{JK} L^K \,.
\label{kvarbc}
\end{align}
This expression tells us that the boundary contribution vanishes if
the spinor superfield satisfies the condition
\begin{align}
L^I = P^{IJ} L^J \,,
\label{scond}
\end{align}
at the worldsheet boundary $\partial \Sigma$, because
$\bar{\eta}^I \Gamma_* \tau_1^{IJ} L^J=
-\bar{\eta}^I \Gamma_* \tau_1^{IJ} P^{JK} L^K=
-\bar{\eta}^I \Gamma_* \tau_1^{IJ} L^J$
means $\bar{\eta}^I \Gamma_* \tau_1^{IJ} L^J=0$.
In what follows, we will show that the spinor superfield indeed
follows the boundary condition (\ref{scond}).

Let us first consider the boundary condition for D$p$-branes with
$p \ge 1$, leaving the discussion of the D-instanton case separately.
In order to see the effect of imposing the boundary condition on
the spinor superfield $L^I$, we focus on the term
of the form $\mathcal{M}^{2n} D\theta$, which is the field
dependent part of the summand in the series expression of
$L^I$ (\ref{superfld}).  For the elementary piece  $\mathcal{M}^2$,
it is not difficult to show that
\begin{align}
(\mathcal{M}^2)^{IJ} = P^{IK} (\mathcal{M}^2)^{KL} P^{LJ}
\end{align}
by using the definition of $\mathcal{M}^2$ given in (\ref{m2dt})
and the boundary condition (\ref{bc}) with (\ref{susybc}). This
means that $(\mathcal{M}^{2n})^{IJ} = P^{IK}
(\mathcal{M}^{2n})^{KL} P^{LJ}$ from the property of $P^{IJ}$
(\ref{bcprop}) and in turn we get $(\mathcal{M}^{2n})^{IJ}
(D\theta)^J = P^{IJ} (\mathcal{M}^{2n})^{JK} (P D \theta)^K$ at
the boundary. Now the question is whether or not the relation
$(D\theta)^I = (P D\theta)^I$ (or $(D P\theta)^I = (P D\theta)^I$)
holds at the boundary. As mentioned in (\ref{diffcov}) in the
previous section, this relation does not hold generically and $(D
P \theta)^I$ differs from $(P D\theta)^I$ by an amount of spin
connection dependent term.  However, what we are interested in
here are the 1/2-BPS D$p$-branes with $p \ge 1$, which should be
located at the coordinate origin in the transverse directions.
Because the spin connection vanishes at such position, we can set
$(D P\theta)^I = (P D\theta)^I$ at least for the description of
1/2-BPS D-branes.  As a result of this, it turns out that the
spinor superfield satisfies the boundary condition (\ref{scond}).
We would like to note that, in showing the boundary condition
(\ref{scond}), we have not imposed any additional boundary
condition other than that introduced in the previous section.
Therefore we conclude that the classification of 1/2-BPS
D$p$-branes $(p \ge 1)$ in the previous section is valid even at
higher orders in $\theta$.

We now turn to the D-instanton case.  This is the special case
in the sense that the sign in (\ref{kvarbc}) is positive instead of
negative basically because the D-instanton is not included in the
criterion given in (\ref{susybc}) and thus the previous argument for
showing the vanishing of the boundary contribution from the
$\kappa$-symmetry transformation is not applicable.
Actually, this is the reason that the
D-instanton should be treated separately.

Let us observe that $D\theta^I = d\theta^I$ at the boundary for
the D-instanton boundary condition because $dX^A=0$ for all $A$.
By using this fact and the D-instanton boundary condition matrix
$P^{IJ} = i \epsilon^{IJ}$ from (\ref{pmat}), we can easily show
that $(\mathcal{M}^2)^{IJ} D \theta^J = (\mathcal{M}^2)^{IJ} d
\theta^J = 0$. This immediately means that the spinor superfield
is given by $L^I = d\theta^I$ at the boundary, which simplifies
the $\kappa$-symmetry transformation (\ref{k-rule}) related to
$L^I$ as $\delta_\kappa \theta = \eta^I$.  Therefore, the boundary
contribution (\ref{kvar}) does not vanish and becomes $2
\int_{\partial \Sigma} \delta_\kappa \bar{\theta}^I \Gamma_*
\tau_1^{IJ} d\theta^J$.  However, this is the term exactly
cancelled by the boundary contribution from the total derivative
term in (\ref{wz}) which has been omitted in this section, and
thus we have $\kappa$-symmetry as a whole.

In conclusion, since we do not have to introduce any additional boundary
condition for showing the $\kappa$-symmetry invariance of the action even
at higher orders in $\theta$, it is verified that the classification of
1/2-BPS D-branes in the previous section is valid at full orders in
$\theta$.

\section{Conclusion}

We have taken the type IIB superstring action in the AdS$_5 {}\times{}$S$^5$
background whose WZ term is of the bilinear form, and given
a covariant open string description of 1/2-BPS D-branes in the background.
Under the $\kappa$-symmetry transformation, some boundary contributions
appear from the variation of the WZ term.  A set of suitable open string
boundary conditions for making them vanish has been investigated to have
the full $\kappa$-symmetry and, as a result, the classification of
possible 1/2-BPS D-branes has been obtained with its validity
check at full orders in $\theta$.  As it should be, our
result agrees exactly with the previous classification
\cite{Skenderis:2002vf,Sakaguchi:2003py}.

The important point in our study is that the bilinear WZ term should
not have the total derivative term for the correct D-brane
classification.  Actually, the description of 1/2-BPS D$p$-branes with
$p \ge 1$ is not sensitive for such term.  However the total derivative term
should be absent for describing the 1/2 BPS D-instanton.  Related
to the present work, there has been an attempt to describe the D-branes
in the AdS$_5 {}\times{}$S$^5$ background from a different perspective,
the integrability, which is based on the same action as ours but keeping
the total derivative term \cite{Dekel:2011ja}.  One of the results was that
D-instanton was not in the class of integrable boundary condition
and excluded in the D-brane classification.  Although our focus
is not the integrability and thus the direct comparison of our result
with that obtained in \cite{Dekel:2011ja} may not be sensible,
one may expect carefully that the absence of the integrable
boundary condition for the supersymmetric D-instanton may be related to
the presence of the total derivative term in the bilinear WZ term.

\section*{Acknowledgments}

This research was supported by Basic Science Research Program through the National Research Foundation of Korea (NRF) funded by the Ministry of Education, Science and Technology with the Grants
No.~2012R1A1A2004203(HS), No.~2009-0084601(HN) and by NRF-2011-0025517(ECY).




\appendix
\section{Supergeometry of the AdS$_5 {}\times{}$S$^5$ background}
\label{app1}

The notation for the supercoordinate we use is
\begin{align}
Z^M = (X^\mu, \theta^I) \,,
\end{align}
where the spinor index for the fermionic coordinate $\theta$ has been
suppressed, $\mu$ is the ten dimensional curved space-time vector index,
and $I$ ($=1,2$) is introduced to distinguish the two same chirality
spinors.
As for the Lorentz frame or the tangent space, the vector index is
denoted by
\begin{align}
A = (a, a') \,, \quad a=0,1,2,3,4 \,, \quad a'=5,\dots,9 \,,
\end{align}
where $a$ ($a'$) corresponds to the tangent space of AdS$_5$ ($S^5$), and
the metric $\eta_{AB}$ follows the most plus sign convention as
$\eta_{AB} = \text{diag} (-, +, +, \dots, +)$.

The matrices acting on the spinors indexed with $I,J,\dots$
are denoted by
\begin{align}
\tau_i^{IJ}\,, \quad (i=1,2,3) \,,
\end{align}
which are the usual Pauli matrices.

The explicit expression for the vector (spinor) superfield or the
Maurer-Cartan one-form superfield
$L^A = dZ^M L^A_M$ ($L^I = dZ^M L_M^I$) is given by
\cite{Metsaev:1998it,Kallosh:1998zx}
\begin{align}
L^A &= e^A + 2 i \sum_{n=0}^{15} \frac{1}{(2n+2)!}
                 \bar{\theta}^I \Gamma^A
                 ({\mathcal M}^{2n})^{IJ} D\theta^J  \,,
\notag \\
L^I &= \sum^{16}_{n=0} \frac{1}{(2n+1)!} (\mathcal{M}^{2n})^{IJ}
      D\theta^J
\,,
\label{superfld}
\end{align}
where ${\mathcal M}^2$ and the spinor covariant derivative $D\theta^I$
are, in the 32 component notation,
\begin{gather}
({\mathcal M}^2)^{IJ}
 = - \epsilon^{IK} \Gamma_* \Gamma^A \theta^K \bar{\theta}^J \Gamma_A
   + \frac{1}{2} \epsilon^{JK}
       ( \Gamma^{ab} \theta^I \bar{\theta}^K \Gamma_{ab} \Gamma_*
        -\Gamma^{a'b'} \theta^I \bar{\theta}^K \Gamma_{a'b'} \Gamma_*') \,,
\notag \\
D\theta^I
= \left( d + \frac{1}{4} \omega^{AB}\Gamma_{AB} \right) \theta^I
   - \frac{i}{2} \epsilon^{IJ} e^A \Gamma_* \Gamma_A \theta^J \,.
\label{m2dt}
\end{gather}
with the convention $\epsilon^{12}=1$ for the antisymmetric tensor
$\epsilon^{IJ}$ $(= i \tau_2^{IJ})$.
Some definitions of gamma matrix products and their properties are
as follows.\footnote{We follow the notation adopted in
\cite{Callan:2003xr}.}
\begin{gather}
\Gamma_* \equiv i \Gamma_{01234} \,, \quad
\Gamma_*' \equiv i \Gamma_{56789} \,, \quad
\Gamma_*^2 = 1 \,, \quad \Gamma_*'^2=-1 \,,
\notag \\
\Gamma^{11} = \Gamma^{01\dots 9} = \Gamma_* \Gamma_*' \,, \quad
(\Gamma^{11})^2 = 1 \,.
\label{gdef}
\end{gather}

The zehnbein and the corresponding spin connection for the
AdS$_5 {}\times{}$S$^5$ are given by \cite{Hatsuda:2002hz}
\begin{gather}
e^a = dX^a + \left( \frac{\sinh X}{X}-1 \right) dX^b Y_b^a \,, \quad
e^{a'} = dX^{a'} + \left( \frac{\sinh X'}{X'}-1 \right)
               dX^{b'} Y_{b'}^{a'} \,,
\notag \\
\omega^{ab} = \frac{1}{2} \left( \frac{\sinh (X/2) }{X/2} \right)^2
               dX^{[a} X^{b]} \,, \quad
\omega^{a'b'} = -\frac{1}{2} \left( \frac{\sinh (X'/2) }{X'/2} \right)^2
               dX^{[a'} X^{b']} \,,
\label{zehn}
\end{gather}
where
\begin{gather}
X = \sqrt{ X^a X_a } \,, \quad X' = \sqrt{ X^{a'} X_{a'} } \,,
\notag \\
Y_a^b = \delta_a^b - \frac{X_a X^b}{X^2} \,, \quad
Y_{a'}^{b'} = \delta_{a'}^{b'} - \frac{X_{a'} X^{b'}}{X'^2} \,.
\end{gather}


\begin{thebibliography}{10}

\bibitem{AdS/CFT}
J.~M.~Maldacena,
``The large N limit of superconformal field theories and supergravity,''
Adv.\ Theor.\ Math.\ Phys.\  {\bf 2} (1998) 231
[Int.\ J.\ Theor.\ Phys.\  {\bf 38} (1999) 1113] [arXiv:hep-th/9711200].

\bibitem{GKPW}
S.~S.~Gubser, I.~R.~Klebanov and A.~M.~Polyakov,
``Gauge theory correlators from non-critical string theory,''
Phys.\ Lett.\ B {\bf 428} (1998) 105 [arXiv:hep-th/9802109]; \\
E.~Witten,
``Anti-de Sitter space and holography,''
Adv.\ Theor.\ Math.\ Phys.\  {\bf 2} (1998) 253 [arXiv:hep-th/9802150].

\bibitem{Metsaev:1998it}
  R.~R.~Metsaev and A.~A.~Tseytlin,
  ``Type IIB superstring action in AdS(5) x S**5 background,''
  Nucl.\ Phys.\ B {\bf 533} (1998) 109
  [hep-th/9805028].

\bibitem{Kallosh:1998zx}
  R.~Kallosh, J.~Rahmfeld and A.~Rajaraman,
  ``Near horizon superspace,''
  JHEP {\bf 9809} (1998) 002
  [arXiv:hep-th/9805217].

\bibitem{Berkovits:1999zq}
  N.~Berkovits, M.~Bershadsky, T.~Hauer, S.~Zhukov and B.~Zwiebach,
  ``Superstring theory on AdS(2) x S**2 as a coset supermanifold,''
  Nucl.\ Phys.\ B {\bf 567} (2000) 61
  [hep-th/9907200].

\bibitem{Roiban:2000yy}
  R.~Roiban and W.~Siegel,
  ``Superstrings on AdS(5) x S**5 supertwistor space,''
  JHEP {\bf 0011} (2000) 024
  [hep-th/0010104].

\bibitem{Bena:2003wd}
  I.~Bena, J.~Polchinski and R.~Roiban,
  ``Hidden symmetries of the AdS(5) x S**5 superstring,''
  Phys.\ Rev.\ D {\bf 69} (2004) 046002
  [hep-th/0305116].

\bibitem{Arutyunov:2009ga}
  G.~Arutyunov and S.~Frolov,
  ``Foundations of the AdS$_5$ x S$^5$ Superstring. Part I,''
  J.\ Phys.\ A A {\bf 42} (2009) 254003
  [arXiv:0901.4937 [hep-th]].

\bibitem{Hatsuda:2000mn}
  M.~Hatsuda, K.~Kamimura and M.~Sakaguchi,
  ``Nondegenerate super anti-de Sitter algebra and a superstring action,''
  Phys.\ Rev.\ D {\bf 62} (2000) 105024
  [hep-th/0007009].

\bibitem{Hatsuda:2001pp}
  M.~Hatsuda and M.~Sakaguchi,
  ``Wess-Zumino term for the AdS superstring and generalized Inonu-Wigner
   contraction,''
  Prog.\ Theor.\ Phys.\  {\bf 109} (2003) 853
  [hep-th/0106114].

\bibitem{Hatsuda:2001xf}
  M.~Hatsuda and K.~Kamimura,
  ``Classical AdS superstring mechanics,''
  Nucl.\ Phys.\ B {\bf 611} (2001) 77
  [hep-th/0106202].

\bibitem{Polyakov:2004br}
  A.~M.~Polyakov,
  ``Conformal fixed points of unidentified gauge theories,''
  Mod.\ Phys.\ Lett.\ A {\bf 19} (2004) 1649
  [hep-th/0405106].

\bibitem{Hatsuda:2002hz}
  M.~Hatsuda and M.~Sakaguchi,
  ``Wess-Zumino term for AdS superstring,''
  Phys.\ Rev.\ D {\bf 66} (2002) 045020
  [hep-th/0205092].

\bibitem{Bain:2002tq}
  P.~Bain, K.~Peeters and M.~Zamaklar,
  ``D-branes in a plane wave from covariant open strings,''
  Phys.\ Rev.\ D {\bf 67} (2003) 066001
  [hep-th/0208038].

\bibitem{Skenderis:2002vf}
  K.~Skenderis and M.~Taylor,
  ``Branes in AdS and pp wave space-times,''
  JHEP {\bf 0206} (2002) 025
  [hep-th/0204054].

\bibitem{Sakaguchi:2003py}
  M.~Sakaguchi and K.~Yoshida,
  ``D-branes of covariant AdS superstrings,''
  Nucl.\ Phys.\ B {\bf 684} (2004) 100
  [hep-th/0310228].

\bibitem{Lambert:1999id}
  N.~D.~Lambert and P.~C.~West,
  ``D-branes in the Green-Schwarz formalism,''
  Phys.\ Lett.\ B {\bf 459} (1999) 515
  [hep-th/9905031].

\bibitem{Hyun:2002xe}
  S.~-j.~Hyun, J.~Park and H.~-j.~Shin,
  ``Covariant description of D-branes in IIA plane wave background,''
  Phys.\ Lett.\ B {\bf 559} (2003) 80
  [hep-th/0212343].

\bibitem{Hatsuda:2004vi}
  M.~Hatsuda and K.~Kamimura,
  ``Wess-Zumino terms for AdS D-branes,''
  Nucl.\ Phys.\ B {\bf 703} (2004) 277
  [hep-th/0405202].

\bibitem{Sakaguchi:2004}
  M.~Sakaguchi and K.~Yoshida,
  ``Notes on D-branes of type IIB string on AdS(5) x S**5,''
  Phys.\ Lett.\ B {\bf 591} (2004) 318
  [hep-th/0403243].

\bibitem{Dekel:2011ja}
  A.~Dekel and Y.~Oz,
  ``Integrability of Green-Schwarz Sigma Models with Boundaries,''
  JHEP {\bf 1108} (2011) 004
  [arXiv:1106.3446 [hep-th]].




\bibitem{Callan:2003xr}
  C.~G.~Callan, Jr., H.~K.~Lee, T.~McLoughlin, J.~H.~Schwarz, I.~Swanson
  and X.~Wu,
  ``Quantizing string theory in AdS(5) x S**5: Beyond the pp wave,''
  Nucl.\ Phys.\ B {\bf 673} (2003) 3
  [hep-th/0307032].




\end{thebibliography}
\end{document}